\documentclass[twocolumn,twoside]{revtex4}
\usepackage{graphicx}
\usepackage{fancyhdr}
\usepackage{amsmath, amssymb}
\usepackage{color}
\begin{document}

\title{First row CKM unitarity}

%

\author{Chien-Yeah Seng}
\affiliation{$^{b}$Helmholtz-Institut f\"ur Strahlen- und Kernphysik 
	and Bethe Center for Theoretical Physics,\\ Universit\"at Bonn, 
	53115 Bonn, Germany}

\begin{abstract}
In this talk I briefly review the precision test of the first row CKM unitarity, focusing mainly on $V_{ud}$ and $V_{us}$ extracted from pion, kaon, neutron and superallowed nuclear beta decays. I will discuss the current status of several important Standard Model theory inputs to these processes, and the need for future improvements. 
\end{abstract}

\maketitle

\thispagestyle{fancy}


\section{Introduction}
This article serves a a contribution to the Proceedings of the 20$^{\textrm{th}}$ Conference on Flavor Physics and 
CP Violation (FPCP~2022) based on a remote talk I gave. Most of the contents can be found in a recent review paper~\cite{Seng:2021gmh}, with only several minor updates. 

Despite being one of the most successful physics theory ever, the Standard Model (SM) of particle physics falls short in explaining some cosmological observations such as dark matter, dark energy and matter-antimatter asymmetry, and does not address a number of theory-driven questions including the hierarchy problem and the unification of forces. This leads to world-wide programs to search for physics beyond the Standard Model (BSM) in all energy scales, and the precision test of the first row Cabibbo-Kobayashi-Maskawa (CKM) matrix~\cite{Cabibbo:1963yz,Kobayashi:1973fv} unitarity, $|V_{ud}|^2+|V_{us}|^2+|V_{ub}|^2=1$, represents one of the most powerful tools for that purpose at low energies.
Due to the smallness of $|V_{ub}|^2$, at the current precision level it reduces to the Cabibbo unitarity $|V_{ud}|^2+|V_{us}|^2=1$, so the test of the unitarity relation is equivalent to checking the mutual consistencies in the extraction of the Cabibbo angle $\theta_C=\cos^{-1}|V_{ud}|=\sin^{-1}|V_{us}|$ from different experiments. 

\begin{figure}[t]
	\begin{centering}
		\includegraphics[width=0.8\linewidth]{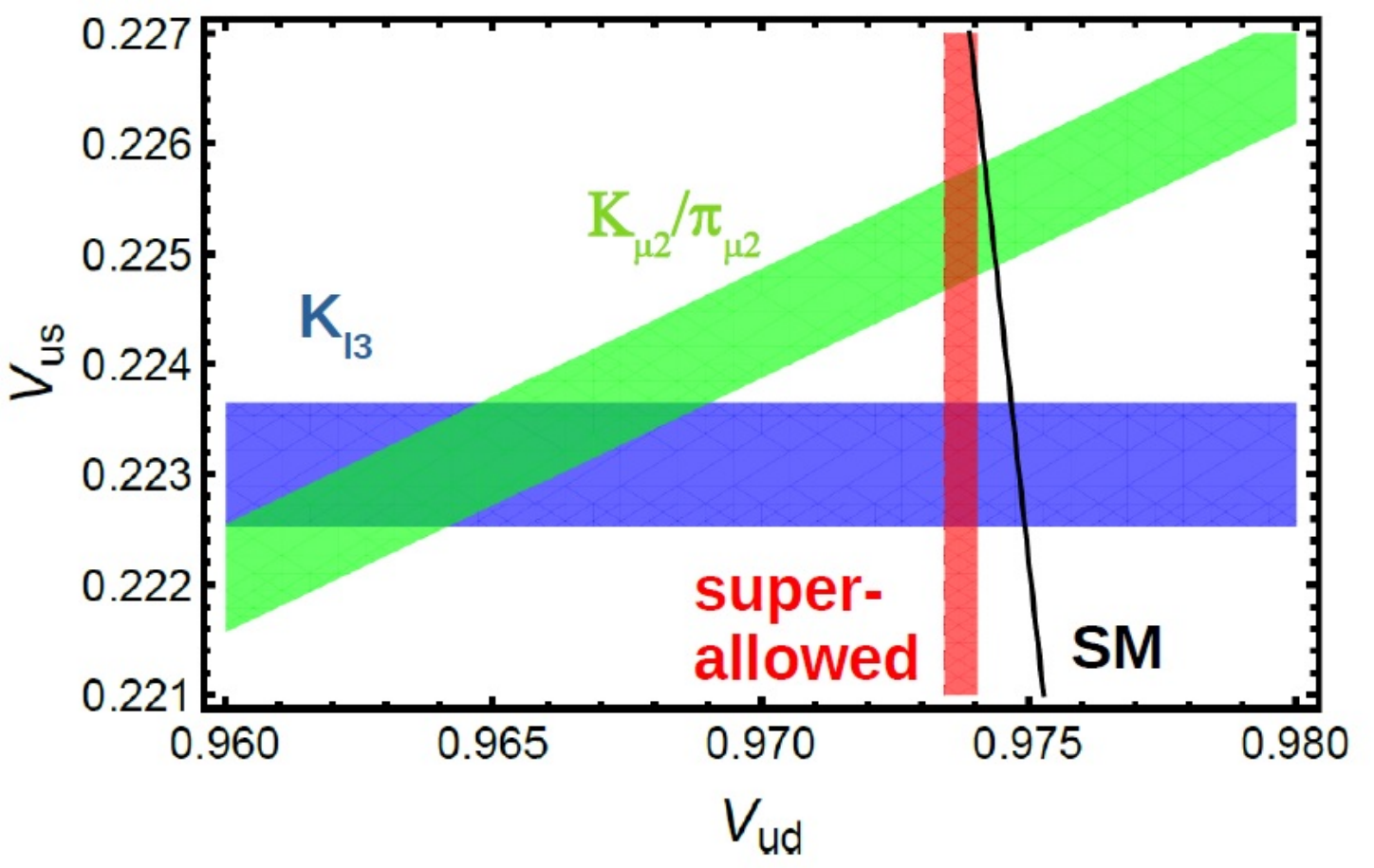}
		\par\end{centering}
	\caption{
		The current status of $|V_{ud}|$ and $|V_{us}|$ obtained from superallowed nuclear beta decays (red band), semileptonic kaon decays (blue band) and leptonic kaon/pion decays (green band). The black line represents $|V_{ud}|^2+|V_{us}|^2=1$, i.e. the first row CKM unitarity requirement. 
		\label{fig:VudVus}
	}
\end{figure}

Due to the limitation of time, I will concentrate mainly on decays of pion, kaon, neutron and $J^P=0^+$ nuclei. The values of $|V_{ud}|$ and $|V_{us}|$ extracted from these decay processes are summarized in Fig.\ref{fig:VudVus}. If different experiments are consistent with each other and satisfy the SM prediction, then there should be a common overlapping region between all the colored bands and the black line; but clearly, such region does not exist. From the figure we observe several interesting anomalies: For instance, the distance between the blue + red region and the black line signifies the breaking of the first row CKM unitarity by combining $|V_{ud}|$ from superallowed nuclear decays (i.e. beta decays of $0^+$ nuclei) and $|V_{us}|$ from semileptonic kaon decays ($K_{\ell 3}$)~\cite{Seng:2021nar}:
\begin{equation}
|V_{ud}|_{0^+}^2+|V_{us}|_{K_{\ell 3}}^2-1=-0.0021(7)~.
\end{equation}
Meanwhile, the distance between the red + blue region and the red + green region may be interpreted as an inconsistency in the measurement of $|V_{us}|$ from $K_{\ell 3}$ and leptonic kaon decays ($K_{\mu 2}$)~\cite{Seng:2022wcw}:
\begin{equation}
|V_{us}|=\left\{ \begin{array}{ccc}
0.22308(55) &  & K_{\ell3}\\
0.2252(5) &  & K_{\mu2}
\end{array}\right.~.
\end{equation}
Both anomalies are at the level of $3\sigma$, and provide interesting hints for BSM physics~\cite{Crivellin:2020lzu,Crivellin:2021njn}.

\begin{table}
	\begin{centering}
		\begin{tabular}{|c|c|}
			\hline 
			$|V_{ud}|_{0^{+}}^{2}+|V_{us}|_{K_{\ell3}}^{2}-1$ & $-2.1\times10^{-3}$\tabularnewline
			\hline 
			\hline 
			$\delta|V_{ud}|_{0^{+}}^{2}$, exp & $2.1\times10^{-4}$\tabularnewline
			\hline 
			$\delta|V_{ud}|_{0^{+}}^{2}$, RC & $1.8\times10^{-4}$\tabularnewline
			\hline 
			$\delta|V_{ud}|_{0^{+}}^{2}$, NS & $5.3\times10^{-4}$\tabularnewline
			\hline 
			$\delta|V_{us}|_{K_{\ell3}}^{2}$, exp + th & $1.8\times10^{-4}$\tabularnewline
			\hline 
			$\delta|V_{us}|_{K_{\ell3}}^{2}$, lat & $1.7\times10^{-4}$\tabularnewline
			\hline 
			Total uncertainty & $6.5\times10^{-4}$\tabularnewline
			\hline 
			Significance level & 3.2$\sigma$\tabularnewline
			\hline 
		\end{tabular}
		\par\end{centering}
	\caption{Error budget of $|V_{ud}|_{0^+}^2$ and $|V_{us}|_{K_{\ell 3}}^2$.\label{tab:split}}
	
\end{table}

The conclusions above are based on a combination of experimental measurements and SM theory inputs. To get a feeling, let us take a closer look to the unitarity violation with $|V_{ud}|_{0^+}$ and $|V_{us}|_{K_{\ell 3}}$, which is summarized in Table~\ref{tab:split}. The current significance level is $3.2\sigma$, and there are five major sources of uncertainty prohibiting us from claiming a discovery:
\begin{enumerate}
	\item $\delta|V_{ud}|^2_{0^+}$, exp: The experimental errors in the superallowed decay half life;
	\item  $\delta|V_{ud}|^2_{0^+}$, RC: Theory errors from the single-nucleon radiative corrections (RC) in free neutron and nuclear beta decay;
	\item  $\delta|V_{ud}|^2_{0^+}$, NS: Theory errors from the nuclear-structure-dependent corrections in superallowed decays;
	\item  $\delta|V_{us}|^2_{K_{\ell3}}$, exp + th: The combined experimental + theory (non-lattice) errors in $K_{\ell 3}$, and
	\item $\delta|V_{us}|^2_{K_{\ell3}}$, lat: The lattice errors in the $K\rightarrow\pi$ transition form factor at zero momentum transfer.
\end{enumerate}
In the $V_{ud}$ determination, the dominant uncertainty comes from theory, whereas in $V_{us}$ the experimental and theory uncertainties are comparable. In this talk I will focus more on the theory ones.

\section{Inputs in the nucleon/nuclear sector}

\begin{figure}[t]
	\begin{centering}
		\includegraphics[width=0.8\linewidth]{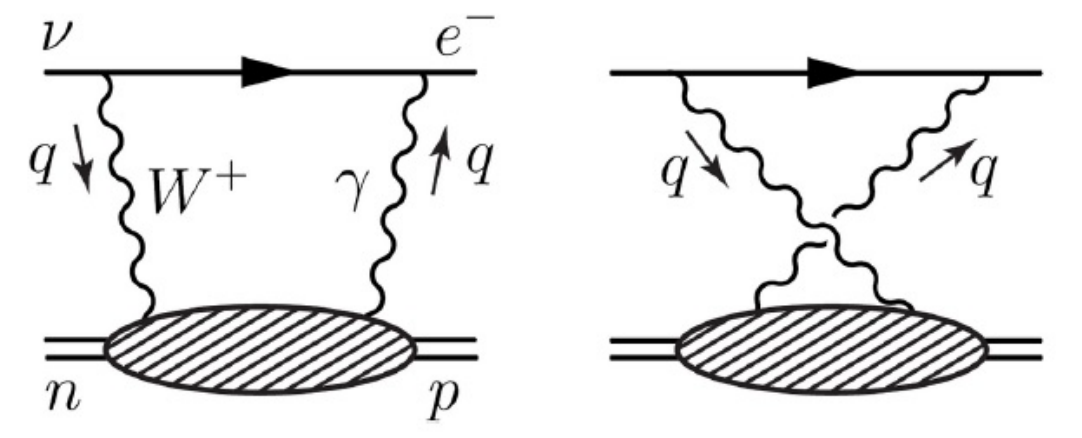}
		\par\end{centering}
	\caption{The single-nucleon $\gamma W$-box diagram.
		\label{fig:gammaW}
	}
\end{figure}

Beta decays of free neutron and nuclear systems are primary avenues to obtain $|V_{ud}|$, and we may start by discussing the theory inputs in the single nucleon sector. The major source of theory uncertainty is the so-called $\gamma W$-box diagram, where the nucleon exchanges a $W$-boson and a photon with the lepton (see Fig.\ref{fig:gammaW}). Its precise determination is challenging because the loop integral probes all momentum scales from infrared to ultraviolet. In particular, at $Q\sim 1$~GeV the strong interaction governed by Quantum Chromodynamics (QCD) becomes non-perturbative and results in a large hadronic uncertainty, which represents a major theory challenge in the past 4 decades~\cite{Sirlin:1977sv}. The best treatment before 2018 consists of dividing the loop integral into different regions according to $Q^2$; at large $Q^2$ perturbative QCD is applicable, at small $Q^2$ the dominance of the elastic contribution is assumed, while at intermediate $Q^2$ an interpolating function is constructed to connect the high and low $Q^2$ result~\cite{Marciano:2005ec}.

In year 2018 a dispersion relation (DR) treatment~\cite{Seng:2018yzq,Seng:2018qru} was introduced to relate the loop integral to experimentally-measurable structure functions~\footnote{The normalization of $F_3^{(0)}$ in different literature may differ by a factor 2.}:
\begin{equation}
\Box_{\gamma W}=\frac{2\alpha}{\pi}\int_0^\infty\frac{dQ^2}{Q^2}\frac{M_W^2}{M_W^2+Q^2}\int_0^1dx\frac{1+2r}{(1+r)^2}F_3^{(0)}~,
\end{equation}
where $x$ is the Bjorken variable, $r=\sqrt{1+4m_N^2x^2/Q^2}$, and $F_3^{(0)}$ is a parity-odd, spin-independent structure function resulting from the product of an isosinglet vector current and an isotriplet axial current. It could be related, barring some model-dependence, to a similar structure function $F_3^{\nu p+\bar{\nu}p}$ obtained from inclusive $\nu p/\bar{\nu}p$ scattering experiments. Making use of existing data , Refs.\cite{Seng:2018yzq,Seng:2018qru} reduced the theory uncertainty in $\Box_{\gamma W}$ substantially, and at the same time a large shift of the central value of $\Box_{\gamma W}$ was observed. It reduced the value of $|V_{ud}|$ from 0.97420(21) in earlier 2018 to 0.97370(14) in late 2018, and unveiled a tension in the first row CKM unitarity. This finding was later confirmed by several independent studies~\cite{Czarnecki:2019mwq,Seng:2020wjq,Hayen:2020cxh,Shiells:2020fqp}. 

A major limiting factor of the DR treatment is the low quality of the neutrino scattering data in the most interesting region of $Q^2\sim 1$~GeV$^2$. To overcome this limitation, there is an ongoing program to calculate the box diagram directly using lattice QCD. The first attempt towards this direction was done in Ref.\cite{Feng:2020zdc}, where a simpler pion axial $\gamma W$-box diagram was computed from first principles to a 1\% precision by combining 4-loop pQCD prediction at large $Q^2$ and lattice calculations at small $Q^2$. It led to a significant reduction of the theory uncertainty in the pion semileptonic decay ($\pi_{e3}$), making it the theoretically cleanest avenue for the $V_{ud}$ extraction. Also, it provides indirect implications to the nucleon box diagram through a Regge exchange picture~\cite{Seng:2020wjq}. A next step is to compute the nucleon box diagram on lattice; it may proceed in the same way which the pion box diagram is calculated, or with alternative approaches, e.g. the Feynman-Hellmann theorem~\cite{Seng:2019plg}. 

Next we proceed from a free neutron to nuclear systems. Superallowed beta decays of $J^P=0^+$, $I=1$ nuclei currently provide the best measurement of $V_{ud}$, mainly due to two reasons: First, since the nuclei are spinless, at tree level they probe only the conserved vector current which matrix element is completely fixed by isospin symmetry. Second, a large number of superallowed transitions are measured, with 15 among them whose lifetime precision is 0.23\% or better~\cite{Hardy:2020qwl}. This provides a huge gain in statistics.

The advantages above come with a price, namely the nuclear-structure-dependent theory uncertainties. The master formula for the $V_{ud}$ extraction from superallowed decay reads:
\begin{equation}
|V_{ud}|_{0^+}^2=\frac{2984.43~s}{\mathcal{F}t(1+\Delta_R^V)}\label{eq:supermaster}
\end{equation} 
where $\Delta_R^V$ denotes the single-nucleon RC we discussed before. The nuclear-structure-dependent corrections are lumped into the ``corrected'' $ft$-value:
\begin{equation}
\mathcal{F}t=ft(1+\delta_\text{R}')(1+\delta_\text{NS}-\delta_\text{C}).
\end{equation}
There are three types of nucleus-dependent corrections: (1) The ``outer'' correction $\delta_\text{R}'$, (2) $\delta_\text{NS}$ which is the nuclear structure effects in the ``inner'' RC, and (3) The isospin-breaking (ISB) correction $\delta_\text{C}$. The first is well under control so I will focus on the next two. 

$\delta_\text{NS}$ originates from the difference between the nuclear and single-nucleon axial $\gamma W$-box diagram, i.e. 
\begin{equation}
\Box_{\gamma W}^\text{nucl}=\Box_{\gamma W}^n+\left[\Box_{\gamma W}^\text{nucl}-\Box_{\gamma W}^n\right].
\end{equation}
The term in the square bracket gives rise to $\delta_\text{NS}$ after integrating over the phase space. There are two ways that a non-zero difference could occur: (1) The single-nucleon absorption spectrum at low energies is distorted by nuclear corrections, and (2) The two gauge bosons may couple to two distinct nucleons in the nucleus, which does not have a counterpart in the single nucleon sector.
Earlier studies of $\delta_\text{NS}$ made use of a nuclear shell model~\cite{Hardy:2014qxa}; in particular, a quenching factor is used to account for the reduced strength of the Born contribution in the nuclear medium~\cite{Towner:1994mw}. However, it was recently pointed out that such calculations missed some of the very important nuclear effects, for instance the contribution from a quasi-free nucleon in a nucleus~\cite{Seng:2018qru}. It was also argued that, unlike free nucleon, the electron energy-dependence in the nuclear box diagram may be non-negligible~\cite{Gorchtein:2018fxl}. A simple Fermi gas model suggested that these two new nuclear corrections carry different sign and partially cancel each other, resulting in an inflated theory uncertainty in $\delta_\text{NS}$. This leads to the present quoted value of $|V_{ud}|=0.97373(31)$~\cite{Hardy:2020qwl}, where the dominant uncertainty comes from $\delta_\text{NS}$. In the future, direct computations of the nuclear $\gamma W$-box diagram using ab-initio methods are expected to reduce such uncertainty.

Finally, the ISB correction $\delta_\text{C}$, which mainly originates from the Coulomb interaction between protons, modifies the tree-level charged weak matrix element from the group theory prediction of $\sqrt{2}$. This correction is important in terms of aligning the $\mathcal{F}t$ values of different superallowed transitions, i.e. to ensure that the right hand side of Eq.\eqref{eq:supermaster} is nucleus-independent. Again, this correction has been studied systematically by Hardy and Towner using nuclear shell model, which achieved an impressive alignment of the $\mathcal{F}t$ values~\cite{Towner:2002rg,Towner:2007np,Hardy:2008gy}. But at the same time their results are also questioned in several aspects, including theoretical inconsistencies~\cite{Miller:2008my,Miller:2009cg,Condren:2022dji} and not being able to be reproduced by other nuclear theory calculations~\cite{Ormand:1989hm,Ormand:1995df,Satula:2011br,Liang:2009pf,Auerbach:2008ut,Damgaard:1969yyx}, which generically predict smaller values of $\delta_\text{C}$ (that would lead to smaller $|V_{ud}|$ and further intensify the first row unitary violation). In the future, a more model-independent approach is much desirable to pin down this correction.

\section{Inputs in the kaon/pion sector}

Next I will discuss the extraction of $V_{us}$ from kaon decays. In the leptonic decay channel, it is usually the ratio between the kaon and pion leptonic decay rate $R_A\equiv \Gamma_{K_{\mu 2}}/\Gamma_{\pi_{\mu2}}$ that is analyzed~\cite{Marciano:2004uf}, which returns the ratio $|V_{us}/V_{ud}|$:
\begin{equation}
\frac{|V_{us}|f_{K^+}}{|V_{ud}|f_{\pi^+}}=\left[\frac{ M_{\pi^+}}{M_{K^+}}R_A\right]^{1/2}\frac{1-m_\mu^2/M_{\pi^+}^2}{1-m_\mu^2/M_{K^+}^2}(1-\delta_\text{EM}/2).
\end{equation}
The advantage is that several SM corrections common to the kaon and pion channel cancel out in the ratio, making it theoretically cleaner than the individual channels. The two main theory inputs to the formula above are as follows. First, $f_{K^+}/f_{\pi^+}$, namely the ratio between the charged kaon and the charged pion decay constant, is taken from lattice QCD calculations. The global averages with different number of active quark flavors ($N_f$) agree with each other and can be found in the FLAG review~\cite{Aoki:2021kgd}. Here we just quote the most precise average which comes from $N_f=2+1+1$:
\begin{equation}
N_f=2+1+1:f_{K^+}/f_{\pi^+}=1.1932(21)~\text{\cite{Bazavov:2017lyh,Dowdall:2013rya,Carrasco:2014poa,Miller:2020xhy}}~.
\end{equation}
The second input is the residual electromagnetic RC which does not fully cancel between the numerator and the denominator. Chiral Perturbation Theory  (ChPT) provides a solid prediction of this quantity at leading order because it is independent from poorly-constrained low energy constants (LECs)~\cite{Knecht:1999ag,Cirigliano:2011tm}:
\begin{equation}
\delta_\text{EM}=\delta_\text{EM}^K-\delta_\text{EM}^\pi=-0.0069(17).
\end{equation}
This result is recently confirmed by a direct lattice QCD calculation~\cite{Giusti:2017dwk}, which serves as a strong proof of the reliability of the theory input. With them we obtain:
\begin{equation}
|V_{us}/V_{ud}|=0.23131(41)_\text{lat}(24)_\text{exp}(19)_\text{RC}.
\end{equation}
Usually, this is later combined with $|V_{ud}|$ obtained from superallowed decays to obtain the value of $|V_{us}|$, i.e. the green + red region in Fig.\ref{fig:VudVus}.

Next we proceed to $K_{\ell 3}$, which is another best avenue for the precision measurement of $V_{us}$. There are six independent channels for this decay, which correspond to $K_L$, $K_S$ and $K^+$ decaying into $e^+$ and $\mu^+$ respectively, and measurements of branching ratio exist in all channels~\cite{Zyla:2020zbs}. To extract $|V_{us}|$ one makes use of the following master formula:
\begin{eqnarray}
\Gamma_{K_{\ell 3}}&=&\frac{G_F^2|V_{us}|^2M_K^5C_K^2}{192\pi^3}S_\text{EW}|f_+(0)|^2I_{K\ell}^{(0)}\nonumber\\
&&\times\left(1+\delta_\text{EM}^{K\ell}+\delta_\text{SU(2)}^{K\pi}\right).
\end{eqnarray}
I will now discuss the theory inputs that appear in the formula above. First, $C_K$ is a trivial isospin factor that equals 1 for $K_{\ell 3}^0$ and $1/\sqrt{2}$ for $K_{\ell 3}^+$. Next, $S_\text{EW}$ is a channel-independent multiplicative factor that encodes the short-distance electroweak RC. Of course how it separates from the ``long-distance'' part is merely a choice, and for standard analysis one always use the value $S_\text{EW}=1.0232(3)_\text{HO}$~\cite{Marciano:1993sh}. While these two are not an issue, the truly non-trivial theory inputs are as follows. 

First, $|f_+(0)|$ is the $K^0\rightarrow\pi^-$ transition form factor at zero momentum transfer (this channel is chosen just by convention). In the $\text{SU(3)}_f$ limit it is exactly 1, but the fact that $m_s\neq \hat{m}$ brings upon a small correction. This constant is computed with lattice QCD, and here we quote the FLAG average at different $N_f$~\cite{Aoki:2021kgd}:
\begin{eqnarray}
N_f=2+1+1&:&f_+(0)=0.9698(17)~\text{\cite{Carrasco:2016kpy,Bazavov:2018kjg}}\nonumber\\
N_f=2+1&:&f_+(0)=0.9677(27)~\text{\cite{Bazavov:2012cd,Boyle:2015hfa}}\nonumber\\
N_f=2&:&f_+(0)=0.9560(57)(62)~\text{\cite{Lubicz:2009ht}}~.\nonumber\\
\end{eqnarray}
Usually only the $N_f=2+1+1$ and $N_f=2+1$ results are used for the $|V_{us}|$ determination. But even these are not without ambiguity: For instance, a recent calculation from the PACS collaboration at $N_f=2+1$ with two lattice spacings gives $f_+(0)=0.9615(10)(^{+47}_{-2})(5)$, significantly smaller than the FLAG average ~\cite{Ishikawa:2022otj}. If this value is used, then there will be no observable difference between the $K_{\ell 3}$ and $K_{\mu 2}$ determinations of $|V_{us}|$. It is therefore important to resolve the discrepancy between different lattice results and make sure that there is no large hidden systematic errors in those calculations.

\begin{table}
	\begin{centering}
		\begin{tabular}{|c|c|}
			\hline 
			& $I_{K\ell}^{(0)}$\tabularnewline
			\hline 
			\hline 
			$K^0e$ & 0.15470(15)\tabularnewline
			\hline 
			$K^+e$ & 0.15915(15)\tabularnewline
			\hline 
			$K^0\mu$ & 0.10247(15)\tabularnewline
			\hline 
			$K^+\mu$ & 0.10553(16)\tabularnewline
			\hline 
		\end{tabular}
		\par\end{centering}
	\caption{$K_{\ell3}$ phase space factors from the dispersive parameterization.\label{tab:PS}}
	
\end{table}

Next we need the phase-space factor
\begin{eqnarray}
I_{K\ell}^{(0)}&=&\int_{m_\ell^2}^{(M_K-M_\pi)^2}\frac{dt}{M_K^8}\bar{\lambda}^{3/2}\left(1+\frac{m_\ell^2}{2t}\right)\left(1-\frac{m_\ell^2}{t}\right)^2\nonumber\\
&&\times\left[\bar{f}_+^2(t)+\frac{3m_\ell^2\Delta_{K\pi}^2}{(2t+m_\ell^2)\bar{\lambda}}\bar{f}_0^2(t)\right]
\end{eqnarray}
(where $\bar{\lambda}=[t-(M_K+M_\pi)^2][t-(M_K-M_\pi)^2]$ and $\Delta_{K\pi}=M_K^2-M_\pi^2$)
which probes the $t$-dependence of the rescaled $K\pi$ form factors $\bar{f}_+(t)$ and $\bar{f}_0(t)$. It is usually obtained by fitting to the $K_{\ell 3}$ Dalitz plot with specific parameterizations of the form factors. Among them, the dispersive parameterization currently quotes the smallest uncertainty (at 0.1\% level) and is usually taken as the standard input~\cite{PSCKM21}; we quote their results in Table~\ref{tab:PS}. 

\begin{figure}[t]
	\begin{centering}
		\includegraphics[width=1.0\linewidth]{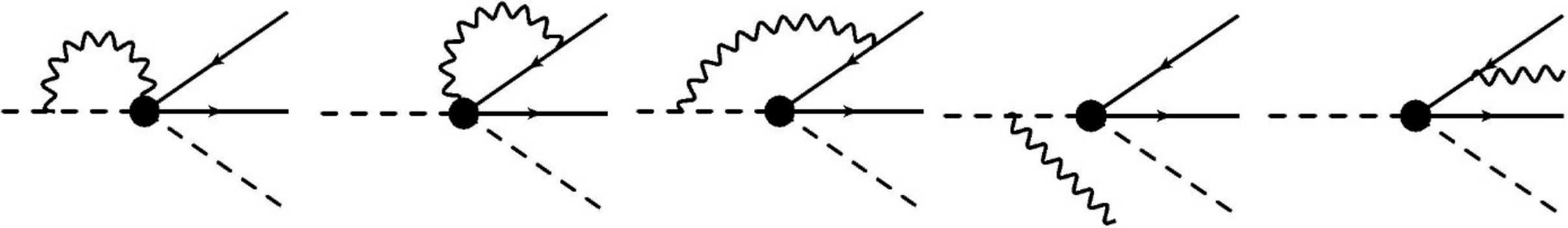}
		\par\end{centering}
	\caption{Long-distance electromagnetic RC in $K_{\ell3}$, self-energy diagrams are not shown.
		\label{fig:EMRC}
	}
\end{figure}

\begin{table}
	\begin{centering}
		\begin{tabular}{|c|c|c|}
			\hline 
			& Sirlin's representation & ChPT\tabularnewline
			\hline 
			\hline 
			$K^{0}e$ & $11.6(2)(1)(1)(2)$ & $9.9(1.9)(1.1)$\tabularnewline
			\hline 
			$K^{+}e$ & $2.1(2)(1)(4)(1)$ & $1.0(1.9)(1.6)$\tabularnewline
			\hline 
			$K^{0}\mu$ & $15.4(2)(1)(1)(2)(2)$ & $14.0(1.9)(1.1)$\tabularnewline
			\hline 
			$K^{+}\mu$ & $0.5(2)(1)(4)(2)(2)$ & $0.2(1.9)(1.6)$\tabularnewline
			\hline 
		\end{tabular}
		\par\end{centering}
	\caption{Results of $\delta_{\text{EM}}^{K\ell}$ in units of $10^{-3}$,
		from Sirlin's representation and ChPT.\label{tab:EMRC}}
	
\end{table}

Next we have $\delta_\text{EM}^{K\ell}$, which denotes the long-distance electromagnetic RC to the decay rate. It consists of virtual photon loop corrections and real bremsstrahlung corrections, see Fig.\ref{fig:EMRC}. The previous state-of-the-art calculation of this correction comes from ChPT~\cite{Cirigliano:2008wn}, where the two major sources of uncertainties are (1) The discarded terms of higher chiral power counting, and (2) The unknown LECs. Recently a series of re-analysis of $\delta_\text{EM}^{K\ell}$ is performed~\cite{Seng:2021boy,Seng:2021wcf,Seng:2022wcw} based on a combination of Sirlin's representation of the electroweak RC~\cite{Sirlin:1977sv,Seng:2021syx} and ChPT~\cite{Seng:2019lxf}, which allows a resummation of the most important terms to all orders in the chiral power counting; in addition, the use of most recent lattice QCD inputs effectively pinned down the LECs~\cite{Seng:2020jtz,Ma:2021azh}. While agreeing with the ChPT result, this re-analysis substantially reduces the theory uncertainty by nearly an order of magnitude, reaching the level of $10^{-4}$. We compare the outcomes of these two determinations in Table~\ref{tab:EMRC}.

Finally, the ISB correction
\begin{equation}
\delta_\text{SU(2)}^{K^+\pi^0}\equiv \left(\frac{f_+^{K^+\pi^0}(0)}{f_+^{K^0\pi^-}(0)}\right)^2-1
\end{equation}
measures the difference of the $K\rightarrow \pi$ transition form factor between the $K^+$ and $K^0$ channel. Upon neglecting small electromagnetic contributions, it is expressed in terms of the quark mass parameters $m_s$ and $\hat{m}$~\cite{Antonelli:2010yf}:
\begin{equation}
\delta_\text{SU(2)}^{K^+\pi^0}=\frac{3}{2}\frac{1}{Q^2}\left[\frac{\hat{M}_K^2}{\hat{M}_\pi^2}+\frac{\chi_{p^4}}{2}\left(1+\frac{m_s}{\hat{m}}\right)\right]
\end{equation}
where $\hat{M}_{K,\pi}$ are the meson masses in the isospin limit, $Q^2=(m_s^2-\hat{m}^2)/(m_d^2-m_u^2)$, and $\chi_{p^4}$ is a calculable coefficient. The most recent lattice QCD inputs give $Q=23.3(5)$ and $m_s/\hat{m}=27.42(12)$ at $N_f=2+1$~\cite{Blum:2014tka,Durr:2010vn,Durr:2010aw,MILC:2009ltw,Fodor:2016bgu}, which return $\delta_\text{SU(2)}^{K^+\pi^0}=0.0457(20)$. On the other hand, phenomenological inputs based on $\eta\rightarrow 3\pi$ returns a somewhat larger value of $\delta_\text{SU(2)}^{K^+\pi^0}=0.0522(34)$~\cite{Colangelo:2018jxw}; the discrepancy between these two determinations is not yet fully understood. In what follows we adopt the lattice determination.

\begin{table}
	\begin{centering}
		\begin{tabular}{|c|c|}
			\hline 
			& $|V_{us}f_{+}(0)|$\tabularnewline
			\hline 
			\hline 
			$K_{L}e$ & $0.21617(46)_{\text{exp}}(10)_{I_{K}}(4)_{\text{EM}}(3)_\text{HO}$\tabularnewline
			\hline 
			$K_{S}e$ & $0.21530(122)_{\text{exp}}(10)_{I_{K}}(4)_{\text{EM}}(3)_\text{HO}$\tabularnewline
			\hline 
			$K^{+}e$ & $0.21714(88)_{\text{exp}}(10)_{I_{K}}(21)_{\text{ISB}}(5)_{\text{EM}}(3)_\text{HO}$\tabularnewline
			\hline 
			$K_{L}\mu$ & $0.21649(50)_{\text{exp}}(16)_{I_{K}}(4)_{\text{EM}}(3)_\text{HO}$\tabularnewline
			\hline 
			$K_{S}\mu$ & $0.21251(466)_{\text{exp}}(16)_{I_{K}}(4)_{\text{EM}}(3)_\text{HO}$\tabularnewline
			\hline 
			$K^{+}\mu$ & $0.21699(108)_{\text{exp}}(16)_{I_{K}}(21)_{\text{ISB}}(6)_{\text{EM}}(3)_\text{HO}$\tabularnewline
			\hline 
			Average: $Ke$ & $0.21626(40)_{K}(3)_{\text{HO}}$\tabularnewline
			\hline 
			Average: $K\mu$ & $0.21654(48)_{K}(3)_{\text{HO}}$\tabularnewline
			\hline 
			Average: tot & $0.21634(38)_{K}(3)_{\text{HO}}$\tabularnewline
			\hline 
		\end{tabular}
		\par\end{centering}
	\caption{Channel-dependent and averaged values of $|V_{us}f_{+}(0)|$.\label{tab:Vusfp}}
	
\end{table}

With the inputs above, we may determine the value of $|V_{us}f_+(0)|$ from each of the six channels as well as averages over different channels~\cite{Seng:2021nar,Seng:2022wcw}, which we summarize in Table~\ref{tab:Vusfp}. The averages over the $Ke$ and $K\mu$ channels agree with each other within respective uncertainties, and do not show signatures of lepton flavor non-universality. Supplementing with the $N_f=2+1+1$ lattice average of $f_+(0)$ gives $|V_{us}|_{K_{\ell 3}}=0.22308(39)_\text{lat}(39)_K(3)_\text{HO}$.
From Table~\ref{tab:Vusfp} it appears that experimental uncertainties dominate over the non-lattice theory uncertainties, but one still needs to further scrutinize all the theory inputs to make sure that the present $V_{us}$ anomaly does not come from some unexpected, large SM corrections.

Finally, I will briefly discuss a relatively new idea to determine $V_{us}/V_{ud}$. Ref.\cite{Czarnecki:2019iwz} suggested that, in addition to the axial ratio $R_A=\Gamma(K_{\mu 2})/\Gamma(\pi_{\mu 2})$, one may also use the vector ratio $R_V\equiv \Gamma(K_{\ell 3})/\Gamma(\pi_{e 3})$ as an alternative avenue to determine the ratio $V_{us}/V_{ud}$, which may shed new lights on  the $V_{us}$ anomaly. The short-distance SM corrections as well as some possible BSM corrections that affect $K_{\ell 3}$ and $\pi_{e 3}$ in the same way cancel out in the ratio, which limits the possible explanations if the anomaly persists.

The cancellation of the long-distance electromagnetic RC in $R_V$ is not as good as in $R_A$ (for example, the $\mathcal{O}(e^2p^2)$ LECs do not fully cancel out), but this is not an issue anymore due to the recent improvements of the $K_{\ell 3}$ and $\pi_{e3}$ RC we described above. As a consequence, $R_V$ is theoretically cleaner than $R_A$, as can be seen from the following formula:
\begin{eqnarray}
\left|\frac{V_{us}f_+^K(0)}{V_{ud}f_+^\pi(0)}\right|&=&0.22216(64)_{\text{BR}(\pi_{e3})}(39)_K(2)_{\tau_{\pi^+}}(1)_{\text{RC}_\pi}
\nonumber\\
\left|\frac{V_{us}f_{K^+}}{V_{ud}f_{\pi^+}}\right|&=&0.27600(29)_\text{exp}(23)_\text{RC}.
\end{eqnarray}
The first line is from $R_V$ and the second line from $R_A$. The major limiting factor, however, comes from experiments, in particular the large uncertainty in the $\pi_{e3}$ branching ratio. The current best measurement comes from the PIBETA experiment in year 2004~\cite{Pocanic:2003pf,Czarnecki:2019iwz}:
\begin{equation}
\text{BR}(\pi_{e3})=1.038(6)\times 10^{-8},
\end{equation}
so there are a lot of rooms for improvement. In fact, partially motivated by the new idea of $R_V$ and the theory progress of the RC, a next-generation experiment (PIONEER) of rare pion decays is aiming to improve the BR($\pi_{e3}$) precision by a factor of 3 or more~\cite{PIONEER:2022yag,PIONEER:2022alm}. This will make $R_V$ competitive to $R_A$ in the determination of $V_{us}/V_{ud}$.

\section{Summary}

To summarize, I described several anomalies at the level of $3\sigma$ that have been observed in the measurements of the first row CKM matrix element $V_{ud}$ and $V_{us}$ from various beta decay processes. They provide interesting hints to new physics so it is important to confirm them with higher precision by improving the SM theory inputs. In the $V_{ud}$ sector, we need to improve the RC in the single-nucleon and nuclear systems using lattice QCD and ab-initio methods respectively, and need a more model-independent determination of the ISB corrections in the nuclear wavefunctions. In the $V_{us}$ sector, we need better lattice inputs for the kaon/pion decay constants and the $K\pi$ transition form factor, and further improvements of the RC in leptonic and semileptonic kaon decays. The phase space factor in $K_{\ell 3}$ and the ISB correction to the $K^+_{\ell 3}$ channels also need to be better understood. Successful reduction of all these theory uncertainties, say by a factor 1/2, could increase the significance of the anomalies to more than 5$\sigma$ assuming the central values remain unchanged. Of course there are also many desirable experimental improvements, for example better measurements of the $K_{\ell 3}$ and $\pi_{e3}$ branching ratios, the neutron lifetime, and the neutron axial coupling constant $g_A$. 

\begin{acknowledgments}
I thank Xu Feng, Daniel Galviz, Mikhail Gorchtein, Lu-Chang Jin, Peng-Xiang Ma, William J. Marciano, Ulf-G. Mei{\ss}ner, Hiren H. Patel and Michael J. Ramsey-Musolf for collaborations in related topics. 	
This work is supported in
part by the Deutsche Forschungsgemeinschaft (DFG, German Research
Foundation) and the NSFC through the funds provided to the Sino-German Collaborative Research Center TRR110 ``Symmetries and the Emergence of Structure in QCD'' (DFG Project-ID 196253076 - TRR 110, NSFC Grant No. 12070131001).
\end{acknowledgments}

\bigskip 

\bibliography{Ke3_ref}

\end{document}